%% file: weighted.tex
\documentclass[a4paper,12pt]{article}

\usepackage{graphicx}
\usepackage[font=small]{caption}
\usepackage{amsmath,amssymb}
\usepackage{amsthm}
\usepackage{longtable}

\newcommand{\Lag}{\mathcal{L}}
\newcommand{\Cmax}{C_\text{max}}
\newcommand{\Frac}[2]{\displaystyle%
                  \frac{\displaystyle{#1}}{\displaystyle{#2}}}
\newcommand{\Ord}[1]{{\cal O}(#1)}
\newcommand{\Binom}[2]{\left(\hspace{-0.3em}\begin{array}{c}%
                       #1\\#2\end{array}\hspace{-0.3em}\right)}
\newcommand{\Newpage}{}
\newcommand{\TableFontSize}{\footnotesize}

\begin{document}
\title{Counting the number of Feynman Graphs in QCD}
\author{T. Kaneko
\vspace{0.5em} \\
\textsl{Computing Research Center,} \\
\textsl{High Energy Accelerator Research Organization (KEK),} \\
\textsl{1-1 Oho, Tsukuba, Ibaraki 305-0801 Japan}
}
\date{2018.11.12}
\maketitle
\abstract{
Information about the number of Feynman graphs for a given physical 
process in a given field theory is especially useful for confirming 
the result of a Feynman graph generator used in an automatic system 
of perturbative calculations.
A method of counting the number of Feynman graphs 
with weight of symmetry factor
was established based on zero-dimensional field theory, 
and was used in scalar theories and QED.
In this article this method is generalized to more complicated
models by direct calculation of
generating functions on a computer algebra system.
This method is applied to QCD with and without counter terms,
where many higher order are being calculated automatically.
}

\section{Introduction}
A program of generating Feynman graphs is one of the fundamental
components of an automatic calculating system of perturbative 
calculations in field theories.
Such programs have been developed for QED \cite{sasaki, grand} 
and for general field theories \cite{nogueira, grc, feynart, fdc}.
The number of generated graphs of a physical process increases rapidly 
as a function of the number of loops and the number of external
particles.
Although one has to check correctness of the set of generated 
Feynman graphs, it includes too many to perform by hands.
It is desirable to have information about number of Feynman graphs 
calculated independently of graph generation methods.

Usually counting numbers of Feynman graphs have been limited to QED
or scalar theories with one or two kinds of interactions.
The aim of this article is to extend the method to various models,
especially to QCD where many higher order corrections are calculated.

There are two methods known to count the number of Feynman graphs.
One is a combinatorial method in graph theory \cite{enum, nogueira} and
the other is perturbative calculation in 0-dimensional field theory
\cite{cvitanovic,itzykson}.
With the former method, the numbers of graphs in $\phi^3 + \phi^4$ model 
are calculable for general and connected graphs.
However, it seems not easy to extend to more complicated physical models,
where many kinds of particles appear
with many kinds of interactions.

The method based on 0-dimensional field theory,
developed by F. Cvitanovi\'{c} et al. \cite{cvitanovic},
calculates not the plain numbers of Feynman graphs
but weighted numbers of Feynman graphs;
the weight is defined as symmetry factor $1/S$, where $S$ is the order
of the automorphism group of the graph.
This symmetry factor appears in the Feynman rules and
is needed for the calculation of Feynman amplitudes.
With this weight the counted numbers are no more integers but 
rational numbers.
These weighted numbers provide more severe test for a Feynman 
graph generator than the plain numbers of graphs, 
since the symmetry factors are not trivial to calculate.
As this method is based on the formulation of field theories,
it has flexibility in applying to many varieties of physical models.

In Ref.\cite{cvitanovic} 
they constructed recursive relations among Green's functions
based on Dyson-Schwinger equation prepared for model by
model.
The size of the calculation became manageable with this
recursion relation.
However, we can now use powerful computer hardware with computer 
algebra systems.
With these tools one will be able to calculate full generating 
functional of Green's functions directly for more complicated 
physical models.
We have tried this approach and obtained the weighted numbers 
of Feynman graphs in QCD with and without counter terms.

In Section 2, a brief description of the framework 
of the calculation is presented.
Our method is introduced in Section \ref{sec:method}
along with Appendix \ref{a:ppandpt},
and is applied to
some models in Section \ref{sec:models}.
The resulting numbers are shown in Appendix \ref{a:wnumber}.
We discuss recursion relations among Green's functions
in Appendix \ref{a:relation} 
as a generalization of the method 
used in \cite{cvitanovic} based on Dyson-Schwinger equation.
Summary and discussions are given in the Section 5.

\Newpage
\section{Framework of the calculation}

Let $F$ be a field theory in which we calculate Feynman amplitudes
of physical processes.
We consider another field theory $F_0$ in 0-dimensional space-time,
in which each Feynman amplitude of a Feynman graph is 1 except
for coupling constants and symmetry factor.
Summing them up over Feynman graphs in $F_0$, we obtain the number
of Feynman graphs weighted by symmetry factor.
It is noted that $F_0$ is not equal to $F$ in the limit of
0-dimensional space-time.
In Ref.\cite{cvitanovic} the correspondence
between $F$ and $F_0$ was shown.
According to them we summarize the 0-dimensional
field theory and how to construct $F_0$ from $F$.
\begin{enumerate}
\item
All quantities do not depend on space-time coordinate nor on momenta,
because the coordinate itself disappears here.
Especially differentials and integrals over coordinate or momentum space
disappear.

\item
Phases of coupling constants and fields are changed such that 
the factors of power of $i=\sqrt{-1}$ disappear in the action.

\item
Sign factors related to anti-commuting fields are changed to 1.
This implies that fermions and ghosts in $F$ are treated 
as bosons in $F_0$.

\item 
Propagators in $F_0$ are to be equal to 1,
irrespective of spins of particles, internal symmetries 
such as color, 
and whether original particles 
are massless or massive in $F$
(some models with internal symmetries are considered in \cite{cvitanovic}.)
Since the number of graphs is independent of spins and commutation
rules, these characters are dropped too and all fields become scalar
fields, either.
Thus only neutral and charged scalar bosons appear in $F_0$.

\item
Coupling constants are kept for perturbative calculations,
and interactions are defined so that the factors
for vertices in Feynman rules become equal to coupling constant 
without numerical factor.
For example, $\phi^3$ interaction appears as $g/3! \phi^3$
in the Lagrangian of $F_0$.

\end{enumerate}
For example, the Lagrangian of $F_0$ for QCD with $u$- and $d$-quarks is
given by
\begin{equation} \label{l:qcd2}
\begin{split}
\mathcal{L} &
= \frac{1}{2} A^2 
  + \phi_u^{*} \phi_u + \phi_d^{*} \phi_d
  + \phi_g^{*} \phi_g
\\ & \quad
  - \frac{g}{3!} A^3 - \frac{g^2}{4!} A^4 
  - g \phi_u^{*} \phi_u A - g \phi_d^{*} \phi_d A
  - g \phi_g^{*} \phi_g A
,
\end{split}
\end{equation}
where $A$ is neutral scalar field corresponding to gluon,
and $\phi_u$, $\phi_d$ and $\phi_g$ are charged scalar fields
corresponding to $u$-quark, $d$-quark and ghost, respectively.

Let us first consider a model consists of one self-interacting
neutral scalar field $\phi$.
Generating functional $Z[J]$ with source field $J$
is defined in $F$ by:
\begin{align}
Z[J] &= \int [d\phi] e^{i S}
,\\
i S &= i \int  \Bigl\{ \Lag(\phi(x)) + \phi(x) J(x) \Bigr\} \;d x
,\\
\Lag(\phi) &= \frac{1}{2} \phi^2 - V(\phi)
.
\end{align}
It is written after path integral over $\phi$ as:
\begin{align}
\begin{split}
Z[J] &= \frac{Z_1[J]}{Z_1[0]}
,\\
Z_0[J] &= 
 \exp\Bigl[ - \Frac{i}{2} \int J(x) \Delta_F(x-y) J(x)\;dx\,dy\Bigr]
,\\
Z_1[J] &=
 \exp\Bigl[ i \int
   \mathcal{L}_{\text{int}}\Bigl(\Frac{1}{i}\frac{\delta~}{\delta J(z)}\Bigr)
  \Bigr]
  Z_0[J]
.
\end{split}
\end{align}
Green's functions are calculated by:
\begin{equation}
\begin{split}
\tau(x_1, ..., x_n) &
= \Bigl(\frac{1}{i}\Bigr)^n \frac{\delta^n Z[J]}{\delta J(x_n) \cdots J(x_1)}
  \Bigr|_{J=0}
= \langle 0| T(\phi(x_1)\cdots \phi(x_n))|0\rangle
.
\end{split}
\end{equation}

In order to make the contribution of each Feynman graph to 1 
in $F_0$,
we adjust phases of fields and coupling constants such that
\begin{align} \label{s:phi}
i S &
\rightarrow
- S 
= - \frac{1}{2} \phi^2 + V(\phi) - \phi J
.
\end{align}
We have:
\begin{align}
Z[J] &= \int [d\phi] e^{-S}
= \frac{Z_1[J]}{Z_1[0]}
,
\end{align}
where
\begin{align}
Z_1[J] = \exp\Bigl[ V(\phi) \Bigr] Z_0[J]
,\qquad
Z_0[J] = \exp\Bigl[ \Frac{1}{2} J^2 \Bigr]
.
\end{align}
Here $J$ is no more function on coordinate space but a simple variable 
and $\phi$ is a differential operator:
\begin{align}
\phi = \frac{d~}{d J}
.
\end{align}
Generating `functional' becomes a function.

Potential $V(\phi)$ in the case of $\phi^3 + \phi^4$ model with 
coupling constant $g$ becomes
\begin{align} \label{v:phi}
V(\phi) = \frac{g}{3!} \phi^3 + \frac{g^2}{4!} \phi^4
.
\end{align}
In order also to calculate the number of vacuum graphs, we use
$Z_1[J]$ instead of $Z[J]$.
We calculate $Z_1[J]$ directly as a power series with respect
to $g$, with the aid of computer algebra systems,
in limiting by the maximum order of coupling constants $\Cmax$.
It implies that $g^{\Cmax+1}$ in expressions is set to 0.
Thus $Z_1[J]$ becomes a polynomial with respect to the variable 
$g$ of maximum degree $\Cmax$.
Once $Z_1[J]$ is obtained, 
the weighted numbers of connected graphs and 
one-particle-irreducible (1PI)
graphs are calculated in the usual procedure.

\Newpage
\section{Method}
\label{sec:method}

For the general case, we consider arbitrary neutral and 
charged fields.
As we will see in section \ref{sec:models}, the calculation grows
rapidly with the number of variables, and it is desirable to decrease
the number of variables if possible.
When particle numbers of charged fields are conserved,
these fields always appear 
through products of pairs like $\psi^{*} \psi$.
For such fields calculation is accelerated when they are handled
in pairs rather than two independent variables 
$\psi$ and $\psi^{*}$.
Including these situations, we take the following notation 
for representing field
variables:
\begin{center}
\begin{tabular}{lcccl}
Type & Field & Source field & Indices \\
Neutral & $\phi_\mu$ & $J_\mu$ & $\mu=1,...,N_n$, \\
Charged & $\psi_\nu^{*}$ & $\eta_\nu$ & $\nu=1,...,N_c$, \\
     & $\psi_\nu$ & $\eta_\nu^{*}$ \\
Pair & $\rho_\xi = \psi_\xi'^{*} \psi_\xi'$ 
  & $\sigma_\xi = \eta_\xi'^{*}\eta_\xi'$ & $\xi=1,...,N_p$. \\
\end{tabular}
\end{center}

As the action of the model in 0-dimensional space-time, we assume
\begin{equation}
\begin{split}
S &= \frac{1}{2} \sum_\mu \phi_\mu^2 
    + \sum_\nu \psi_\nu^{*} \psi_\nu
    + \sum_\xi \rho_\xi + S_\text{int}(\phi, \psi^{*}, \psi, \rho)
\\ & \quad
    + \sum_\mu \phi_\mu J_\mu 
    + \sum_\nu (\psi_\nu^{*} \eta_\nu + \eta_\nu^{*} \psi_\nu)
    + \sum_\xi (\psi_\xi'^{*} \eta_\xi' + \eta_\xi'^{*} \psi_\xi')
.
\end{split}
\end{equation}

Generating function $Z_1$ of Feynman amplitudes, 
including vacuum graphs, is expressed by:
\begin{align} \label{z1}
Z_1 &= \exp(- S) = \exp(- S_\text{int}) Z_0
,\\
Z_0 &= \exp \Bigl(
          \frac{1}{2} \sum_\mu J_\mu^2 
        + \sum_\nu \eta_\nu^{*} \eta_\nu
        + \sum_\xi \sigma_\xi \Bigr)
,
\end{align}
The field variables in $S_\text{int}$ are replaced by differential
operators:
\begin{gather}
\phi_\mu = \frac{\partial~}{\partial J_\mu}
,\\
\psi_\nu = \frac{\partial~}{\partial \eta_\nu^{*}}
,\qquad
\psi_\nu^{*} = \frac{\partial~}{\partial \eta_\nu}
,\\
\rho_\xi = \frac{\partial~}{\partial \eta_\xi'}
           \frac{\partial~}{\partial \eta_\xi'^{*}}
         =
           \frac{\partial~}{\partial \sigma_\xi}
         + \sigma_\xi  \frac{\partial^2~}{\partial \sigma_\xi^{2}}
.
\end{gather}
We define the following function $Q$ under the rule of 
$g^{\Cmax+1}=0$:
\begin{gather}
Q := Z_0^{-1} Z_1 - 1 
   = Z_0^{-1} \sum_{n=1}^{\Cmax} \frac{1}{n!} (-S_\text{int})^n Z_0
,
\end{gather}
This function is of at least $\Ord{g}$.
We use vector notation of indices:
\begin{equation}
\begin{split}
i &= (i_1, ..., i_\mu, ..., i_{N_n})
,\\
j &= (j_1, ..., j_\nu, ..., j_{N_c})
,\\
k &= (k_1, ..., k_\nu, ..., k_{N_c})
,\\
l &= (l_1, ..., l_\xi, ..., l_{N_p})
,
\end{split}
\end{equation}
Function $Q$ is expressed by:
\begin{align}
Q = Z_0^{-1} \Bigl[
   \sum_{i,j,k,l} a_{i j k l} 
         \prod_{\mu} \phi_\mu^{i_\mu}
         \prod_{\nu} \Bigl\{(\psi_\nu^{*})^{j_\nu}
                     (\psi_\nu)^{k_\nu} \Big\}
         \prod_{\xi} (\rho_\xi)^{l_{\xi}}
     \Bigr] Z_0 - 1
,
\end{align}
where the coefficients $a_{i j k l}$ 
are functions of $g$ and are $\Ord{g}$ except for $a_{0000}=1$,
and fields represent differential operators acting on $Z_0$.
We define the set of functions $P$ and $\tilde{P}$ such that:
\begin{align} \label{def:pm}
\frac{\partial^m~}{\partial J^m} e^{J^2/2} &= P_m(J) e^{J^2/2}
,\\ \label{def:ptmn}
\frac{\partial^m~}{\partial \eta^m} 
\frac{\partial^n~}{\partial \eta^{*n}} 
e^{\eta^{*}\eta} &
= \tilde{P}_{m,n}(\eta^{*}, \eta)e^{\eta^{*}\eta}
,\\ \label{def:ptm}
\Bigl(\frac{\partial~}{\partial \eta'} 
\frac{\partial~}{\partial \eta'^{*}} \Bigr)^m
e^{\eta'^{*}\eta'} &
= \tilde{P}_{m}(\eta'^{*}\eta') e^{\eta'^{*}\eta'}
= \tilde{P}_{m}(\sigma) e^\sigma
.
\end{align}
These functions are polynomials with respect to the source fields.
Concrete forms and formulas
of $P$ and $\tilde{P}$ are given in Appendix \ref{a:ppandpt}.
With these polynomials we obtain
\begin{align} \label{q:pm}
Q = 
    \sum_{i,j,k,l} a_{ijkl} 
         \prod_{\mu} P_{i_\mu}(J_\mu)
         \prod_{\nu} \tilde{P}_{j_\nu k_\nu}(\eta_\nu^{*}, \eta_\nu)
         \prod_{\xi} \tilde{P}_{l_\xi}(\sigma_\xi)
    -1
.
\end{align}
Thus we obtain $Z_1[J]$ explicitly and
the weighted number of Feynman graphs at the same time 
for all possible physical processes 
under the limitation of order of coupling constants by $\Cmax$.

This method,
power series expansion and replacement of the monomials of field variables
by $P$ and $\tilde{P}$, can easily be done with the aid of
a computer algebra system.

The weighted numbers of connected Feynman graphs are obtained
by calculating $\log Z_1$ as usual:
\begin{align}
W_1 := \log Z_1 = \log(1+Q) 
        + \frac{1}{2} \sum_\mu J_\mu^2 
        + \sum_\nu \eta_\nu^{*} \eta_\nu
        + \sum_\xi \sigma_\xi
.
\end{align}
Term $\log(1+Q)$ is calculated by power series expansion:
\begin{align}
\tilde{W} := \log (1+Q) 
  = \sum_{k=1}^{\Cmax} \frac{(-1)^{k+1}}{k} Q^k
.
\end{align}
This calculation is also performed on a computer algebra system.
It is noted that function $\tilde{W}$ is at least $\Ord{g}$.
We calculate $\tilde{W}$ instead of $W_1$;
the differences are 2-point functions of free propagation.
The power series expansion of $\tilde{W}$ is written by:
\begin{align}
\tilde{W} &
= \sum_{ijkl} \frac{c_{ijkl}}{
       \prod_\mu i_\mu! \prod_\nu (j_\nu! k_\nu!) \prod_\xi (l_\xi!)^2}
    \prod_\mu J_\mu^{i_\mu}
    \prod_\nu (\eta_\nu^{*})^{j_\nu} (\eta_\nu)^{k_\nu}
    \prod_\xi \sigma_\xi^{l_\xi}
.
\end{align}
The coefficients $c_{ijkl}$ are the weighted numbers of Feynman graphs
for a process with specified external particles.
In this way we obtain
the weighted numbers of Feynman graphs
for all possible physical processes 
at the same time.

The numbers for 1PI Feynman graphs are calculated
by Legendre transformation as in the usual way:
\begin{equation} \label{gen:gamma0}
\begin{split}
\Gamma_1 &
= W_1 - \sum_\mu J_\mu \phi_\mu
      - \sum_\nu (\eta_\nu^{*} \psi_\nu + \psi_\nu^{*} \eta_\nu)
      - \sum_\xi (\eta_\xi'^{*} \psi_\xi' + \psi_\xi'^{*} \eta_\xi')
\\ &
= \tilde{W} 
      + \sum_\mu \frac{1}{2} (J_\mu - \phi_\mu)^2
      + \sum_\nu (\eta_\nu^{*} - \psi_\nu^{*} ) (\eta_\nu - \psi_\nu )
\\ & \quad
      + \sum_\xi (\eta_\xi'^{*} - \psi_\xi'^{*} ) (\eta_\xi' - \psi_\xi' )
      - \sum_\mu \frac{1}{2} \phi_\mu^2
      - \sum_\nu \psi_\nu^{*} \psi_\nu 
      - \sum_\xi \psi_\xi'^{*} \psi_\xi'
,
\end{split}
\end{equation}
where 
\begin{align}
\phi_\mu  &
= J_\mu + \frac{\partial \tilde{W}}{\partial J_\mu}
,\\
\psi_\nu &
= \eta_\nu + \frac{\partial \tilde{W}}{\partial \eta_\nu^{*}}
,\\
\psi_\nu^{*} &
= \eta_\nu^{*} + \frac{\partial \tilde{W}}{\partial \eta_\nu}
,\\
\psi_\xi' &
= \eta_\xi' + \frac{\partial \tilde{W}}{\partial \eta_\xi'^{*}}
= \eta_\xi' \Bigl(1 + \frac{\partial \tilde{W}}{\partial \sigma_\xi}
           \Bigr)
,\\
\psi_\xi'^{*} &
= \eta_\xi'^{*} + \frac{\partial \tilde{W}}{\partial \eta_\xi'}
= \eta_\xi'^{*} \Bigl(1 + \frac{\partial \tilde{W}}{\partial \sigma_\xi}
           \Bigr)
,\\
\rho_\xi &
= \sigma_\xi \Bigl(1 + \frac{\partial \tilde{W}}{\partial \sigma_\xi}
             \Bigr)^2
,
\end{align}
and
\begin{align}
\eta_\xi'^{*} \psi_\xi' + \psi_\xi'^{*} \eta_\xi' &
= 2 \sigma_\xi 
  \Bigl(1 + \frac{\partial \tilde{W}}{\partial \sigma_\xi} \Bigr)
.
\end{align}
Function $\Gamma_1$ is to be expressed as a function of
$\phi$, $\psi$, $\psi^{*}$ and $\rho = \psi^{*} \psi$,
where source fields $J$, $\eta$, $\eta^*$ and $\sigma$ are eliminated.
We use the following expressions
for the elimination of source fields:
\begin{align}
J_\mu &
= \phi_\mu  - \frac{\partial \tilde{W}}{\partial J_\mu}
,\\
\eta_\nu  &
= \psi_\nu - \frac{\partial \tilde{W}}{\partial \eta_\nu^{*}}
,\\
\eta_\nu^{*} &
= \psi_\nu^{*} - \frac{\partial \tilde{W}}{\partial \eta_\nu}
,\\
\sigma_\xi &
= \rho_\xi - \sigma_\xi \Bigl( 2 + \frac{\partial \tilde{W}}{\partial \sigma_\xi}
           \Bigr) 
           \frac{\partial \tilde{W}}{\partial \sigma_\xi}
.
\end{align}
The right-hand sides of these equations depend on source
fields through derivatives of $\tilde{W}$, which are at least $\Ord{g}$.
We can eliminate source fields on the right-hand sides of these
expressions by repeated substitution of source fields by these
equations themselves.

In order to see how it works, let us consider an example of
a model of one scalar filed.
Since $\tilde{W}$ is a polynomial with respect to source fields $J$
and is $\Ord{g}$, the above equations is written in the following 
form:
\begin{align}
J &
= \phi - \tilde{W}'(J)
= \phi - g (a_1 J + a_2 J^2 + \cdots + a_n J^n)
,
\end{align}
where coefficients $a_j$ are functions of $g$ of $\Ord{g^0}$.
Replacing $J$ on the right hand side by this equation itself, 
we obtain
\begin{align*}
J &
= \phi - \tilde{W}'(\phi - \tilde{W}'(J))
\\ &
= \phi - g [a_1 (\phi - \tilde{W}'(J)) + a_2 (\phi - \tilde{W}'(J))^2 
          + \cdots + a_n (\phi-\tilde{W}'(J))^n]
\\ &
= \phi - \tilde{W}'(\phi)
  + g \tilde{W}'(J) [ a_1 + a_2 (2 \phi - \tilde{W}'(J))
\\ & \qquad
          + \cdots + a_n (n \phi + \cdots + (-\tilde{W}'(J))^{n-1}) ]
.
\end{align*}
Since $\tilde{W}'(J)$ is $\Ord{g}$, the terms dependent on $J$
become of $\Ord{g^2}$.
When this substitution is made once more,
the terms dependent on $J$ become of $\Ord{g^3}$.
Repeating this substitution, at most $\Cmax$ times in this
case, the terms dependent on $J$ become of $\Ord{g^{\Cmax+1}}$
and are set to zero.
Thus we obtain the expression in which $J$ is eliminated.
For general case with $n$ field variables, all source fields are
eliminated by these substitution at most $n^2 \Cmax$ times.
This process is also done on a computer algebra system.

When we put $g \rightarrow 0$ in Eq.(\ref{gen:gamma0}), we obtain
\begin{align} 
\Gamma_1 & \longrightarrow
      - \sum_\mu \frac{1}{2} \phi_\mu^2
      - \sum_\nu \psi_\nu^{*} \psi_\nu 
      - \sum_\xi \psi_\xi'^{*} \psi_\xi'
.
\end{align}
The terms on the right-hand side represent the contribution 
from free propagators to
generating function $\Gamma_1$.
We put aside these terms in our calculation and define 
$\tilde{\Gamma}_1$ by:
\begin{equation} \label{gen:gamma}
\begin{split}
\tilde{\Gamma}_1 &
= \tilde{W} 
  + \sum_\mu \frac{1}{2} \Big(\frac{\partial \tilde{W}}{\partial J}\Bigr)^2
  + \sum_\nu \frac{\partial \tilde{W}}{\partial \eta_\nu}
             \frac{\partial \tilde{W}}{\partial \eta_\nu^{*}}
      + \sum_\xi \sigma
        \Big(\frac{\partial \tilde{W}}{\partial \sigma_\xi}\Bigr)^2
.
\end{split}
\end{equation}

The weighted numbers of 1PI graphs $d_{ijkl}$
are obtained by the power series expansion of $\tilde{\Gamma}_1$
for each process:
\begin{align}
\tilde{\Gamma}_1 &
= \sum_{ijkl} \frac{d_{ijkl}}{
       \prod_\mu i_\mu! \prod_\nu (j_\nu! k_\nu!) \prod_\xi (l_\xi!)^2}
    \prod_\mu \phi_\mu^i 
    \prod_\nu \Bigl\{ (\psi_\nu^{*})^{j_\nu} (\psi_\nu)^{k_\nu} \Bigr\}
    \prod_\xi (\rho_\xi)^{l_\xi}
.
\end{align}

\Newpage
\section{Models}
\label{sec:models}

We have prepared a program written for REDUCE system \cite{reduce}
to implement our algorithm. 
The program is applied to several models.

\subsection{$\phi^3 + \phi^4$ model}

We have calculated the weighted number of graphs in the model 
which consists of one neutral scalar field defined by
(\ref{v:phi}).
The action is:
\begin{align} \label{eq:phi34}
S 
= \frac{1}{2} \phi^2 
  - \frac{g}{3!} \phi^3 - \frac{g^2}{4!} \phi^4
  + \phi J
\end{align}
The resulting numbers of Feynman graphs 
are useful for testing generation of `topologies'
of a graph generation program for QCD, the standard model, and
other renormalizable models.

It took about 9 seconds for $\Cmax = 21$ on a rather old 
Intel machine
with \texttt{Intel(R) Core(TM) i5-2500 CPU}.
A part of resulting numbers is shown in 
Table 1 in Appendix \ref{a:wnumber}.

We have also calculated in the model with counter terms
except tadpoles.
The action is:
\begin{align} \label{eq:phi34ct}
S 
= \frac{1}{2} \phi^2 
  - \Bigl(\sum_{n=0} g^{2n}\Bigr) \Bigl(
     \frac{g^2}{2}\phi^2 + \frac{g}{3!} \phi^3 + \frac{g^2}{4!} \phi^4
    \Bigr)
  + \phi J
.
\end{align}
The resulting numbers are also shown in the same table 
in Appendix \ref{a:wnumber}.
The required CPU time is increased several percent to the 
case of no counter terms.

\subsection{QCD}

We have counted the weighted number of Feynman graphs in QCD
with $N_f$ quark fields.
The action is:
\begin{equation} \label{eq:qcd}
\begin{split}
S &
= \frac{1}{2} A^2 
  + \sum_{j=1}^{N_f} \psi_j'^{*} \psi_j'
  + \psi_g'^{*} \psi_g'
\\ & \quad
  - \frac{g}{3!} A^3 - \frac{g^2}{4!} A^4 
  - g \sum_{j=1}^{N_f} \psi_j'^{*} \psi_j' A
  - g \psi_g'^{*} \psi_g' A
\\ & \quad
  + A J
  + \sum_{j=1}^{N_f} (\psi_j'^{*} \eta_j' + \eta_j'^{*} \psi_j')
  + \psi_g'^{*} \eta_g' + \eta_g'^{*} \psi_g'
,
\end{split}
\end{equation}
where $A$ is a neutral scalar field corresponding to gluon,
$\psi_j'$ is the charged scalar field
corresponding to $j$-th quark, and $\psi_g$ to ghost.
The resulting numbers are shown for the case of $N_f = 6$ in 
Tables 2 in Appendix \ref{a:wnumber}.

\begin{figure}[htb]
\begin{center}
\includegraphics[width=8cm,clip=True,angle=270]{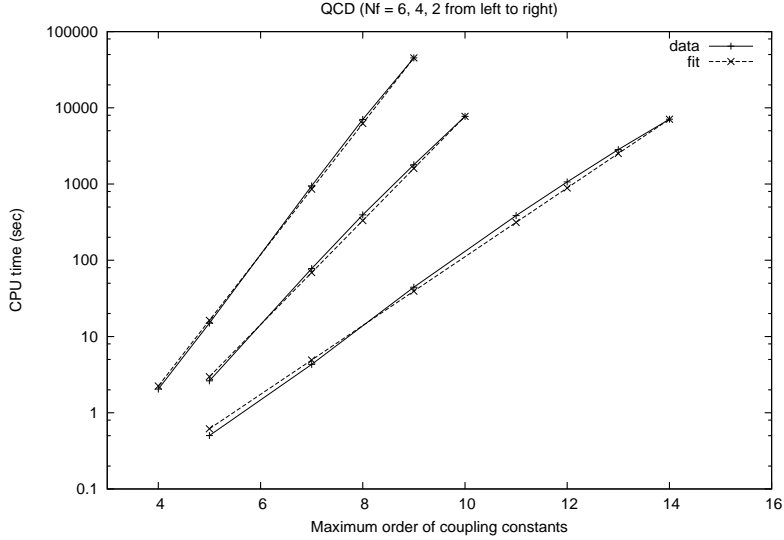}
\end{center}
~\vspace{1em}
\caption{Comparison of CPU time for calculating the number 
of Feynman graphs in QCD with $N_f=$ 2, 4, 6 without
counter terms.
Dashed lines show exponential fit as functions of 
the order of maximum coupling constant
$\Cmax$.
}
\label{fig:cputime}
\end{figure}

We have measured CPU time consumed for $N_f=2$, $4$, $6$
as shown in Fig. \ref{fig:cputime}.
It shows used CPU time grows exponentially for the
order of coupling constants.
The rate of the growth becomes steeper as increasing the number
of fields.

Similar to the case of $\phi^3 + \phi^4$ model, counter terms
are included as shown in the same tables
in Appendix \ref{a:wnumber}.
The action is:
\begin{equation} \label{eq:qcdct}
\begin{split}
S &
= \frac{1}{2} A^2 
  + \sum_{j=1}^{N_f} \psi_j'^{*} \psi_j'
  + \psi_g'^{*} \psi_g'
\\ & \quad
  - \Bigl(\sum_{n=0} g^{2n}\Bigr) \Bigl(
   \frac{g^2}{2} A^2 
  + \sum_{j=1}^{N_f} g^2 \psi_j'^{*} \psi_j'
  + g^2 \psi_g'^{*} \psi_g'
\\ & \qquad
  + \frac{g}{3!} A^3 + \frac{g^2}{4!} A^4 
  + g \sum_{j=1}^{N_f} \psi_j'^{*} \psi_j' A
  + g \psi_g'^{*} \psi_g' A
  \Bigr)
\\ & \quad
  + A J
  + \sum_{j=1}^{N_f} (\psi_j'^{*} \eta_j' + \eta_j'^{*} \psi_j')
  + \psi_g'^{*} \eta_g' + \eta_g'^{*} \psi_g'
.
\end{split}
\end{equation}

\subsection{QED}
For QED, the formulation is to be modified to incorporate
Furry's theorem as described by Ref.\cite{cvitanovic}.
It is done by replacing $Z_1$ of Eq.(\ref{z1}) by
\begin{align*}
Z_1(J, \sigma) &
= (1 - e^2 A^2)^{-1/2}
    \exp\Bigl[ \frac{\sigma}{1 - e A}\Bigr]
     e^{J^2/2}
,
\end{align*}
where $A$ corresponds to photon field, $J$ to its source field,
and $\sigma = \eta'^{*}\eta'$ to the source field of electron-positron
pair, and $e$ is QED coupling constant.
Since electron and positron fields are already integrated out, 
replacement of $\rho^m=(\psi'^{*}\psi')^m$ by $\tilde{P}_m$
is not necessary.
Although the expression of $Q$ alters, the succeeding procedure 
of counting connected and 1PI is the same.
As mentioned in \cite{cvitanovic}, symmetry factors
in QED equal to 1 except for vacuum graphs.
Our program used about 110 seconds for $\Cmax=21$.
Results agreed with Ref.\cite{cvitanovic}.

\Newpage
\section{Summary and Discussions}

A method of calculating the number of Feynman graphs
weighted by symmetry factors for connected and 1PI Feynman graphs
is proposed.
With this method we have calculated in several models, especially in QCD,
with or without counter terms.
This method is to calculate generating functional directly
in 0-dimensional field theory.
The calculation is performed using a computer algebra system.
The main technical points are the replacement of product 
of differential operators by polynomials $P_m$, $\tilde{P}_{m,n}$
and $\tilde{P}_m$,
and calculation of Legendre transformation by repeated substitutions
of variables.

Based on the obtained numbers, systematic testing
tool is prepared in Feynman graph generator \texttt{grc} 
described in Ref.\cite{grc}.

We present some comments in order:
\begin{itemize}

\item 
This method is suitable for automatization.
Once an action of a model is given, calculating procedure
of $Z$, and then the weighted number of connected graphs and 
1PI graphs, is model independent.

\item
As the number of particles increases, the expression 
grows rapidly.
The calculated generating function includes all
possible physical processes only limited by the maximum number of
order of coupling constants.

The total number of processes grows rapidly corresponding to 
the growth of the number of combinations of external particles
of possible processes.
For this reason, the calculation in electro-weak theory, which includes
20 field variables, caused a problem of calculation size.
We have succeeded to calculate up to $\Ord{g^3}$; however,
it failed for higher order calculation in REDUCE system
because of memory problems.

\item
It is possible to count numbers for process by process as
a generalization of the method
described in Ref.\cite{cvitanovic} based on Dyson-Schwinger
equation.
One can write down corresponding recursion relation among
Green's functions based on a differential equation satisfied
by $P_m$ and $\tilde{P}_{m,n}$.
This method is described in Appendix \ref{a:relation}.
We can calculate Green's function one by one for connected
graphs.
However, it will be necessary to follow the same procedure
for Legendre transformation to obtain 1PI graphs.
This method will become
complicated for the models with many particles
such as electro-weak theory.

\item
The most CPU consuming part is one of calculating Legendre
transformation.
Optimization of this part will be depending on 
computer algebra system.

\end{itemize}

\vspace{2em}
\begin{center}
\textbf{Acknowledgments}
\end{center}
The author wish to express his thanks to the members of 
Computing Research Center in KEK
for their support in his scientific activities.
\vspace{2em}
\Newpage
\appendix
\section{Polynomials $P_m$, $\tilde{P}_{m,n}$ and $\tilde{P}_m$}
\label{a:ppandpt}

We look into the polynomial $P_m$, $\tilde{P}_{m,n}$ and 
$\tilde{P}_{m}$ introduced by Eqs.(\ref{def:pm}) -- (\ref{def:ptm}).

\subsection{$P_m$ : neutral scalar fields}

Let $P_m(x)$ be polynomials defined by:
\begin{align}
\label{der:pm}
P_m(x) = e^{-x^2/2} \frac{d^m}{d x^m} e^{x^2/2}
\qquad(j=0,1,2,...)
.
\end{align}
Polynomials $P_m$ can be expressed by Hermite polynomials:%
\footnote{The definition of $H_m$ is different among textbooks.
We use 
$H_n(x) = (-1)^n e^{x^2} \frac{d^n~}{d x^n} e^{-x^2}$
\cite{bateman}.
}
\begin{align}
P_m(x) &
= \Bigl(\frac{i}{\sqrt{2}}\Bigr)^m H_m\Bigl(-\frac{i}{\sqrt{2}}x\Bigr)
.
\end{align}
Generating function is
\begin{align}
e^{t^2/2 + t x} = \sum_{m=0}^\infty P_m(x) \frac{t^m}{m!}
.
\end{align}
The explicit form of $P_m$ is
\begin{align}
P_{2n}(x) &= \sum_{j=0}^n \Binom{n}{j}\frac{(2n-1)!!}{(2j-1)!!} x^{2j}
,\\
P_{2n+1}(x) &= \sum_{j=0}^n \Binom{n}{j} \frac{(2n+1)!!}{(2j+1)!!} x^{2j+1}
.
\end{align}
It is easy to see that $P_m$ satisfies the following relations.
\begin{gather}
P_{m+2}(x) = x P_{m+1}(x) + (m+1) p_m(x)
,\\
\frac{d P_m(x)}{d x} = m P_{m-1}(x)
.
\end{gather}
Each of them satisfies the following differential equation:
\begin{equation} \label{pp:diffeq}
\Bigl(\frac{d^2~}{d x^2} + x \frac{d~}{d x} - m \Bigr) P_m(x)
=0
,
\end{equation}
with initial values at $x=0$:
\begin{equation}
\begin{split} \label{pp:0}
P_{2n}(0) = \frac{(2n)!}{2^n n!}
,& \qquad
P_{2n}'(0) = 0
,\\
P_{2n+1}(0) = 0
,& \qquad
P_{2n+1}'(0) = \frac{(2n+1)!}{2^n n!}
.
\end{split}
\end{equation}

The first several of them are as the following:
\begin{equation}
\begin{split}
P_{0}(x) &= 1
,\\
P_{1}(x) &= x
,\\
P_{2}(x) &= x^2 + 1
,\\
P_{3}(x) &= x^3 + 3 x
,\\
P_{4}(x) &= x^4 + 6 x^2 + 3
,\\
P_{5}(x) &= x^5 + 10 x^3 + 15 x
,\\
P_{6}(x) &= x^6 + 15 x^4 + 45 x^2 + 15
.
\end{split}
\end{equation}
When $P_{j}$ is substituted to Eq.(\ref{q:pm}),
the terms of $P_{j}(x)$ correspond to the same order 
of coupling constants and
the lower degree terms contribute to the processes
with less numbers of external particles.
This implies that these lower degree terms correspond 
to the contiributions from graphs with loops.

\subsection{$\tilde{P}_{m,n}$ : complex scalar fields}

We define $\tilde{P}$ by:
\begin{gather}
\tilde{P}_{m,n}(x, y)
:=
e^{- x y}
\frac{\partial^m~}{\partial x^m} 
\frac{\partial^n~}{\partial y^{n}}
e^{x y}
.
\end{gather}
These polynomials are expressed by Laguerre polynomial $L_n^\alpha$:%
\footnote{
   We use the definition 
$L_n^{\alpha}(x)
= \frac{1}{n!} e^x x^{-\alpha} \frac{d^n~}{d x^n}  e^{-x} x^{n+\alpha}$
.
This is also called Sonin polynomial.
}
\begin{align}
\tilde{P}_{m,n}(x, y) &
= m! \, x^{n-m} L_m^{n-m}(- x y)
= n! \, y^{m-n} L_n^{m-n}(- x y)
.
\end{align}
Generating function of them is
\begin{gather}
e^{s t + x t + y s}
= \sum_{m=0}^\infty \sum_{n=0}^\infty \tilde{P}_{m,n}(x,y)
  \frac{s^m}{m!} \frac{t^n}{n!}
.
\end{gather}
The explicit form is:
\begin{gather}
\tilde{P}_{m,n}(x, y) =
\sum_{k=0}^{\min(m,n)} \frac{m! n!}{k! (m-k)! (n-k)!}
   x^{n-k} y^{m-k}
.
\end{gather}
They satisfy the following recursion relations:
\begin{align}
\tilde{P}_{m,n} (x, y) 
= \frac{\partial \tilde{P}_{m-1,n}}{\partial x}
  + y \tilde{P}_{m-1,n}
\qquad(m \geq 1,\;n \geq 0)
,\\
\tilde{P}_{m,n} (x, y) 
= \frac{\partial \tilde{P}_{m,n-1}}{\partial y}
  + x \tilde{P}_{m,n-1}
\qquad(m \geq 0,\;n \geq 1)
,
\end{align}
and the following differential equations:
\begin{gather} \label{pc:diffeq}
\Bigl(\frac{\partial^2~}{\partial x \partial y}
   + x \frac{\partial~}{\partial x} - n
\Bigr) \tilde{P}_{m,n}(x,y) = 0
,\\ \label{pc:diffeq1}
\Bigl(\frac{\partial^2~}{\partial x \partial y}
   + y \frac{\partial~}{\partial y} - m
\Bigr) \tilde{P}_{m,n}(x,y) = 0
,
\end{gather}
with initial values at the origin:
\begin{equation} \label{pc:0}
\begin{split}
\tilde{P}_{m,n}(0,0) &= m! \delta_{m,n}
,\\
\frac{\partial \tilde{P}_{m,n}}{\partial x}(0,0) = n! \delta_{m+1,n}
,&\qquad
\frac{\partial \tilde{P}_{m,n}}{\partial y}(0,0) = m! \delta_{m,n+1}
.
\end{split}
\end{equation}
The first several of them are shown as the following:
\begin{equation}
\begin{array}{|c|cccc|}
\hline
m\backslash n & 0 & 1 & 2 & 3 \\
\hline
0 & 1 & x & x^2 & x^3 \\
1 & y & 
    xy+1 & 
    x^2y+2x & 
    x^3 y + 3 x^2 \\
2 & y^2 & 
    x y^2 + 2 y & 
    x^2y^2+4 x y + 2 & 
    x^3 y^2 + 6 x^2 y + 6 x \\
3 & y^3 & 
    xy^3 + 3 y^2& 
    x^2 y^3 + 6 x y^2 + 6 y & 
    x^3y^3+9 x^2y^2+18xy + 6 \\
\hline
\end{array}
\end{equation}

\subsection{$\tilde{P}_m$ : pairs of charged fields}

When particle number is conserved for charged particle,
$\tilde{P}_{m,n}$ appear only in the following form,
in which we use $z = x y$:
\begin{align}
\tilde{P}_{m}(z) &:= \tilde{P}_{m,m}(x, y)
= \sum_{k=0}^{m} \frac{1}{k!} \left(\frac{m!}{(m-k)!}\right)^2 z^{m-k}
.
\end{align}
They satisfy the following recursion relation:
\begin{align}
\tilde{P}_{m+1}(z) &=
z \frac{d^2 \tilde{P}_{m}(z)}{d z^2} + (1 + 2 z) \frac{d \tilde{P}_{m}(z)}{d z} 
+ (1+z) \tilde{P}_{m}(z)
,
\end{align}
and differential equation
\begin{equation} \label{pt:diffeq}
\Bigl(x \frac{d^2~}{d x^2} + (x+1) \frac{d~}{d x} - k
    \Bigr) \tilde{P}_k(x)
=0
,
\end{equation}
with initial values at the origin:
\begin{align} \label{pt:0}
\tilde{P}_{m}(0) = m! 
,& \qquad
\tilde{P}_{m}'(0) = m! \, m
.
\end{align}
The first several of them are shown as the following:
\begin{equation}
\begin{split}
\tilde{P}_0(z) &= 1
,\\
\tilde{P}_1(z) &= z + 1
,\\
\tilde{P}_2(z) &= z^{2} + 4 z + 2
,\\
\tilde{P}_3(z) &= z^{3} + 9 z^{2} + 18 z + 6
,\\
\tilde{P}_4(z) &= z^{4} + 16 z^{3} + 72 z^{2} + 96 z + 24
,\\
\tilde{P}_5(z) &= z^{5} + 25 z^{4} + 200 z^{3} + 600 z^{2} + 600 z + 120
,\\
\tilde{P}_6(z) &= z^{6} + 36 z^{5} + 450 z^{4} + 2400 z^{3} + 5400 z^{2} 
               + 4320 z + 720
.
\end{split}
\end{equation}

\Newpage
\section{Recursion relation among Green's functions.}
\label{a:relation}

Let us consider a model consists of gluon and ghost, for simplicity.
The interaction part of action $S_{\text{int}}$ is
\begin{align}
S_{\text{int}} &
= - \frac{g}{3!} (r A)^3 
  - \frac{g^2}{4!} (r A)^4
  - g (s \psi^{*}\psi) (r A)
,
\end{align}
where $A$, $\psi$, and $\psi^{*}$ are scalar fields corresponding
to gluon, ghost, and anti-ghost fields, respectively.
We use variable $\rho$ for the product of $\psi^{*} \psi$,
and $J$ for the source field of $A$ and $\sigma$ for one of $\rho$.
The variables $r$ and $s$ are introduced to count the power
of $A$ and $\rho$, respectively.
The generating function becomes
\begin{gather}
Z_1 
= e^{-S_\text{int}} Z_0 = (1+Q) Z_0
,\qquad
Z_0 
= \exp\Bigl(\frac{1}{2} J^2 + \sigma\Bigr)
,\\
W = \log Z_1
= \log(1 + Q) + \frac{1}{2} J^2 + \sigma
.
\end{gather}
Expanding $e^{-S_\text{int}}$ we obtain
\begin{align}
Q &
= \sum_{l=0} \sum_{m=0} \sum_{n=0}
  \frac{g^{l+2 m + n} r^{3l+4m+n} s^n}{(3!)^l l! (4!)^m m! n!}
  Z_0^{-1} A^{3l+4m+n} \rho^l Z_0 - 1
\\ & \label{zgg:expansion}
= \sum_{l=0} \sum_{m=0} \sum_{n=0}
  \frac{g^{l+2 m + n} r^{3l+4m+n} s^n}{(3!)^l l! (4!)^m m! n!}
  P_{3l+4m+n}(J) \tilde{P}_{n}(\sigma) - 1
.
\end{align}
Using Eqs.(\ref{pp:diffeq}) and (\ref{pt:diffeq})
we obtain the following differential
equations for 
$Q$:
\begin{gather} 
\Bigl(\frac{\partial^2~}{\partial J^2} 
  + J \frac{\partial~}{\partial J} 
  - r \frac{\partial~}{\partial r}
\Bigr) Q = 0
,\\ 
\Bigl(\sigma \frac{\partial^2~}{\partial \sigma^2} 
   + (\sigma+1) \frac{\partial~}{\partial \sigma} 
   - s \frac{\partial~}{\partial s}
\Bigr) Q = 0
.
\end{gather}
From these equations one can obtain differential equations
for $Z_1$.
They correspond to a generalization of Dyson-Schwinger 
equation described in Ref.\cite{cvitanovic}.
We also obtain
differential equations satisfied by the generating function 
$W$ for connected graphs:
\begin{gather} \label{wgg:dj}
      \frac{\partial^2 W}{\partial J^2} 
  + \Bigl(\frac{\partial W}{\partial J}\Bigr)^2
  - J \frac{\partial W}{\partial J} 
  - r \frac{\partial W}{\partial r}
  - 1
 = 0
,\\ \label{wgg:dsigma}
   \sigma \frac{\partial^2 W}{\partial \sigma^2} 
   + \sigma \Bigl(\frac{\partial W}{\partial \sigma} \Bigr)^2
   + (1-\sigma) \frac{\partial W}{\partial \sigma} 
   - s \frac{\partial W}{\partial s}
   - 1
 = 0
.
\end{gather}
Let us define the following functions of $r$ and $s$:
\begin{gather}
W^{(j,k)} 
:= \frac{\partial^{j+k+k} W}{
   \partial J^j \partial \psi^{*k} \partial \psi^k}\Bigr|_{J=\sigma=0}
= k! \frac{\partial^{j+k} W}{
   \partial J^j \partial \sigma^k}\Bigr|_{J=\sigma=0}
.
\end{gather}
They become Green's functions of connected graphs when $r=s=1$.
We obtain recursion relation between $W^{(j,k)}$ from 
the differential equation of $W$:
\begin{align}
W^{(2,0)} &
= \frac{\partial~}{\partial r} W^{(0,0)}
  - (W^{(0,0)})^2 + 1
,\\
W^{(0,1)} &
= \frac{\partial~}{\partial r} W^{(0,0)}
  + 1
,\\
W^{(j+2,0)} &
= \Bigl(r\frac{\partial~}{\partial r}+j\Bigr) W^{(j,0)} 
  - \sum_{l=0}^j \Binom{j}{l} W^{(l+1,0)} \, W^{(j-l+1,0)}
\qquad(j \geq 1)
,\\
\begin{split}
W^{(j,k+1)} &
= \Bigl(s\frac{\partial~}{\partial s}+k\Bigr) W^{(j,k)} 
  - \frac{k}{k+1}\sum_{l=0}^{k-1} 
    \sum_{m=0}^{j}
     \Binom{k-1}{l} 
     \Binom{j}{m}
\\ & \quad\times
     \Binom{k+1}{l+1} 
     W^{(m,l+1)} \, W^{(j-m,k-l)}
\qquad(j \geq 0, \quad k \geq 1)
.
\end{split}
\end{align}
This set of equations implies that
all of $W^{(j,k)}$
can be obtained as functions of $r$ and $s$
with $W^{(0,0)}$ and $W^{(1,0)}$ as inputs.
Let
\begin{align}
Q^{(0,0)} = Q\Bigl|_{J=\sigma=0}
,\qquad
Q^{(1,0)} = \frac{\partial Q}{\partial J}\Bigl|_{J=\sigma=0}
.
\end{align}
They are functions of $r$ and $s$ of $\Ord{g}$ and
are calculable from Eqs.(\ref{zgg:expansion}), (\ref{pp:0}), and
(\ref{pt:0}).
Expanding with respect to the coupling constant $g$,
input functions $W^{(0,0)}$ and $W^{(1,0)}$
are calculable form:
\begin{align}
W^{(0,0)} &
= \log (1+Q^{(0,0)}) = \sum_{l=1} \frac{(-1)^{l+1}}{l} {Q^{(0,0)}}^{l}
,\\
W^{(1,0)} &
= \frac{Q^{(1,0)}}{1+Q^{(0,0)}} = Q^{(1,0)} \sum_{l=0} (-1)^l {Q^{(0,0)}}^l
.
\end{align}
This method makes the problem size smaller, since it enables
to calculate specific processes without calculating all possible
processes.
For 1PI graphs, however, one has to take
the same procedure described in section \ref{sec:method}.

It is straight forward to include other interaction terms such as
counter terms.
In these cases differential equations left unchanged; only
$Q^{(0,0)}$ and $Q^{(1,0)}$ are changed through replacement of
Eq.(\ref{zgg:expansion}).
It is also easy to include quarks by augmenting the number 
of charged scalars and replacing interaction term
$g (s \psi^{*}\psi) (r A)$ by 
$g \sum_i (s_i \psi_i^{*}\psi_i) (r A)$.
For more complicated model, it will be necessary to use differential 
equations for $\tilde{P}_{m,n}(x, y)$ and initial values
given by Eqs.(\ref{pc:diffeq}) -- (\ref{pc:0}).

\Newpage
\section{Calculated numbers of Feynman graphs}
\label{a:wnumber}

Here we show the calculated numbers of Feynman graphs weighted 
by symmetry
factor are shown in the following tables for scalar theory and
QCD.%
\footnote{
More detailed numbers are available from
\texttt{http://research-up.kek.jp/people/kaneko/}.
}

\subsection{Scalar model $\phi^3 + \phi^4$}

The weighted number of Feynman graphs are shown for
$\phi^3+\phi^4$ model in Table 1
for both with and without counter terms described
by Eqs.(\ref{eq:phi34}) and (\ref{eq:phi34ct}).
Column `E' represents the number of external particles
and `L' the number of loops.
Columns with '(CT)' indicate that counter terms are
included in the model.
Calculation was done for $\Cmax = 21$ and the numbers were
obtained from ($\text{E} = 0$, $\text{L} = 11$) to 
($\text{E} = 23$, $\text{L} = 0$).
However we show here a part of them limiting to $\text{E} \leq 6$ and 
$\text{L} \leq 7$.

\newpage
\input{Phi34.tex}
\noindent
\textbf{Table 1} ~
Table of weighted number in
$\phi^3 + \phi^4$ model,
    with and without counter terms for
    (external particles) $\leq 6$ and (loops) $\leq 7$.
    `E' represents the number of external particles
    and `L' the number of loops.
    Symbol `(CT)' indicates that counter terms are included
    in the model.

%
%
%
\subsection{QCD with six quarks}

The weighted number of Feynman graphs is shown for
QCD with six quarks $N_f=6$ in Table 2
for both with and without counter terms.
Column `E' represents the number of external particles,
`L' the number of loops, 
`$gl$', `$q_1$' and `$q_2$' are the number
of external gluons, first and second quarks.
We show only for the case of $q_3=q_4=q_5=q_6=0$.
By renumbering quarks one obtains numbers 
for some other combinations.
It is calculated for $\Cmax = 9$ and the number is
obtained from ($\text{E} = 0$, $\text{L} = 5$) to 
($\text{E} = 11$, $\text{L} = 0$).
However we show here limiting to $\text{E} \leq 4$.

\input{QCD6.tex}
\noindent
\textbf{Table 2} ~ 
Table of weighted number of 
QCD with 6 quarks,
    with and without counter terms.
    `E' represents the number of external particles,
    `L' the number of loops,
    `$gl$' the number of external gluons,
    and `$q_1$', '$q_2$' the number of external quark 1, 2.
    Symbol `(CT)' indicates that counter terms are included
    in the model.

\Newpage

\end{document}

%% file: Phi34.tex
{\TableFontSize
\renewcommand{\arraystretch}{2.2}
\begin{longtable}{|r|r|c|c|c|c|}
\hline
E & L & connected & 1PI & 
        conn. (CT) & 1PI (CT) \\ 
\hline
0 & 2 &   $\Frac{1}{3}$ &   $\Frac{5}{24}$ & 
  $\Frac{5}{6}$ &   $\Frac{17}{24}$ \\ 
0 & 3 &   $\Frac{11}{12}$ &   $\Frac{7}{16}$ & 
  $\Frac{37}{12}$ &   $\Frac{95}{48}$ \\ 
0 & 4 &   $\Frac{1493}{288}$ &   $\Frac{83}{36}$ & 
  $\Frac{5441}{288}$ &   $\Frac{685}{72}$ \\ 
0 & 5 &   $\Frac{25241}{576}$ &   $\Frac{22235}{1152}$ & 
  $\Frac{1345}{8}$ &   $\Frac{86915}{1152}$ \\ 
0 & 6 &   $\Frac{50936927}{103680}$ &   $\Frac{139829}{640}$ & 
  $\Frac{40568905}{20736}$ &   $\Frac{1631659}{1920}$ \\ 
0 & 7 &   $\Frac{284251517}{41472}$ &   $\Frac{64155409}{20736}$ & 
  $\Frac{581444297}{20736}$ &   $\Frac{252924091}{20736}$ \\ 
\hline
E & L & connected & 1PI & 
        conn. (CT) & 1PI (CT) \\ 
\hline
1 & 1 &   $\Frac{1}{2}$ &   $\Frac{1}{2}$ & 
  $\Frac{1}{2}$ &   $\Frac{1}{2}$ \\ 
1 & 2 &   $\Frac{31}{24}$ &   $\Frac{2}{3}$ & 
  $\Frac{67}{24}$ &   $\Frac{5}{3}$ \\ 
1 & 3 &   $\Frac{341}{48}$ &   $\Frac{25}{8}$ & 
  $\Frac{965}{48}$ &   $\Frac{215}{24}$ \\ 
1 & 4 &   $\Frac{22949}{384}$ &   $\Frac{76}{3}$ & 
  $\Frac{73421}{384}$ &   $\Frac{227}{3}$ \\ 
1 & 5 &   $\Frac{1545307}{2304}$ &   $\Frac{41099}{144}$ & 
  $\Frac{5329007}{2304}$ &   $\Frac{128363}{144}$ \\ 
1 & 6 &   $\Frac{777673801}{82944}$ &   $\Frac{194791}{48}$ & 
  $\Frac{2826926677}{82944}$ &   $\Frac{1912765}{144}$ \\ 
1 & 7 &   $\Frac{8665279117}{55296}$ &   $\Frac{119528647}{1728}$ & 
  $\Frac{98228574175}{165888}$ &   $\Frac{408521413}{1728}$ \\ 
\hline
E & L & connected & 1PI & 
        conn. (CT) & 1PI (CT) \\ 
\hline
2 & 1 &   $\Frac{3}{2}$ &   $1$ & 
  $\Frac{5}{2}$ &   $2$ \\ 
2 & 2 &   $\Frac{25}{3}$ &   $\Frac{41}{12}$ & 
  $\Frac{55}{3}$ &   $\Frac{89}{12}$ \\ 
2 & 3 &   $\Frac{1741}{24}$ &   $\Frac{55}{2}$ & 
  $\Frac{4511}{24}$ &   $\Frac{739}{12}$ \\ 
2 & 4 &   $\Frac{80299}{96}$ &   $\Frac{15275}{48}$ & 
  $\Frac{77877}{32}$ &   $\Frac{12445}{16}$ \\ 
2 & 5 &   $\Frac{6869123}{576}$ &   $\Frac{167831}{36}$ & 
  $\Frac{3624397}{96}$ &   $\Frac{1785215}{144}$ \\ 
2 & 6 &   $\Frac{4192377457}{20736}$ &   $\Frac{15668327}{192}$ & 
  $\Frac{14127753307}{20736}$ &   $\Frac{134370073}{576}$ \\ 
2 & 7 &   $\Frac{54906861275}{13824}$ &   $\Frac{713735831}{432}$ & 
  $\Frac{145437685477}{10368}$ &   $\Frac{8671673855}{1728}$ \\ 
\hline
E & L & connected & 1PI & 
        conn. (CT) & 1PI (CT) \\ 
\hline
3 & 0 &   $1$ &   $1$ & 
  $1$ &   $1$ \\ 
3 & 1 &   $\Frac{15}{2}$ &   $\Frac{5}{2}$ & 
  $\Frac{23}{2}$ &   $\Frac{7}{2}$ \\ 
3 & 2 &   $\Frac{1777}{24}$ &   $\Frac{89}{4}$ & 
  $\Frac{3541}{24}$ &   $\Frac{141}{4}$ \\ 
3 & 3 &   $\Frac{44177}{48}$ &   $\Frac{2265}{8}$ & 
  $\Frac{104893}{48}$ &   $\Frac{4207}{8}$ \\ 
3 & 4 &   $\Frac{5292685}{384}$ &   $\Frac{214865}{48}$ & 
  $\Frac{4746983}{128}$ &   $\Frac{454181}{48}$ \\ 
3 & 5 &   $\Frac{556813237}{2304}$ &   $\Frac{665317}{8}$ & 
  $\Frac{1642681493}{2304}$ &   $\Frac{3123995}{16}$ \\ 
3 & 6 &   $\Frac{403304188435}{82944}$ &   $\Frac{112879507}{64}$ & 
  $\Frac{1274880793255}{82944}$ &   $\Frac{288922967}{64}$ \\ 
3 & 7 &   $\Frac{6112916597549}{55296}$ &   $\Frac{72481379183}{1728}$ & 
  $\Frac{20367609554609}{55296}$ &   $\Frac{99633977599}{864}$ \\ 
\hline
E & L & connected & 1PI & 
        conn. (CT) & 1PI (CT) \\ 
\hline
4 & 0 &   $4$ &   $1$ & 
  $4$ &   $1$ \\ 
4 & 1 &   $57$ &   $\Frac{21}{2}$ & 
  $83$ &   $\Frac{23}{2}$ \\ 
4 & 2 &   $\Frac{5057}{6}$ &   $\Frac{709}{4}$ & 
  $\Frac{9431}{6}$ &   $\Frac{977}{4}$ \\ 
4 & 3 &   $\Frac{167621}{12}$ &   $\Frac{26625}{8}$ & 
  $\Frac{93098}{3}$ &   $\Frac{43887}{8}$ \\ 
4 & 4 &   $\Frac{25097635}{96}$ &   $\Frac{1112795}{16}$ & 
  $\Frac{63396679}{96}$ &   $\Frac{2109847}{16}$ \\ 
4 & 5 &   $\Frac{3167606597}{576}$ &   $\Frac{12851447}{8}$ & 
  $\Frac{183533261}{12}$ &   $\Frac{54607215}{16}$ \\ 
4 & 6 &   $\Frac{2675975185651}{20736}$ &   $\Frac{2606881563}{64}$ & 
  $\Frac{8007976663465}{20736}$ &   $\Frac{6084522087}{64}$ \\ 
4 & 7 &   $\Frac{46342382319659}{13824}$ &   $\Frac{72138266309}{64}$ & 
  $\Frac{73377025565557}{6912}$ &   $\Frac{273097894951}{96}$ \\ 
\hline
E & L & connected & 1PI & 
        conn. (CT) & 1PI (CT) \\ 
\hline
5 & 1 &   $\Frac{1149}{2}$ &   $57$ & 
  $\Frac{1609}{2}$ &   $57$ \\ 
5 & 2 &   $\Frac{280735}{24}$ &   $1660$ & 
  $\Frac{497755}{24}$ &   $2110$ \\ 
5 & 3 &   $\Frac{11848865}{48}$ &   $43890$ & 
  $\Frac{8320755}{16}$ &   $66380$ \\ 
5 & 4 &   $\Frac{2154582745}{384}$ &   $1181335$ & 
  $\Frac{1721661675}{128}$ &   $2056895$ \\ 
5 & 5 &   $\Frac{319917889435}{2304}$ &   $\Frac{533959855}{16}$ & 
  $\Frac{846167432375}{2304}$ &   $\Frac{1045018815}{16}$ \\ 
5 & 6 &   $\Frac{310780872308809}{82944}$ &   $\Frac{8002786883}{8}$ & 
  $\Frac{886800014611093}{82944}$ &   $2159061195$ \\ 
5 & 7 &   $\Frac{6083474414980385}{55296}$ &   $31945379025$ & 
  $\Frac{55258819995230675}{165888}$ &   $\Frac{898253508065}{12}$ \\ 
\hline
E & L & connected & 1PI & 
        conn. (CT) & 1PI (CT) \\ 
\hline
6 & 1 &   $7230$ &   $390$ & 
  $9840$ &   $390$ \\ 
6 & 2 &   $\Frac{1149515}{6}$ &   $17865$ & 
  $\Frac{1958885}{6}$ &   $21555$ \\ 
6 & 3 &   $\Frac{59748815}{12}$ &   $641025$ & 
  $10042310$ &   $908550$ \\ 
6 & 4 &   $\Frac{12905505595}{96}$ &   $21758395$ & 
  $\Frac{29598249775}{96}$ &   $35376685$ \\ 
6 & 5 &   $\Frac{2219089049855}{576}$ &   $\Frac{5931760395}{8}$ & 
  $\Frac{312229182035}{32}$ &   $\Frac{10838888715}{8}$ \\ 
6 & 6 &   $\Frac{2449771948135225}{20736}$ &   $\Frac{208155635205}{8}$ & 
  $\Frac{6703936630258195}{20736}$ &   $52505971410$ \\ 
6 & 7 &   $\Frac{53708019970406075}{13824}$ &   $\Frac{7615451146345}{8}$ & 
  $\Frac{117185488235187485}{10368}$ &   $\Frac{8360116793335}{4}$ \\ 
\hline
\end{longtable}
}

%% file: QCD6.tex
{\TableFontSize
\renewcommand{\arraystretch}{2.2}
\begin{longtable}{|r|r|r|r|r|c|c|c|c|}
\hline
E & L & $gl$ & $q_1$ & $q_2$ & connected & 1PI & 
        conn. (CT) & 1PI (CT) \\ 
\hline
0 & 2 & 0 & 0 & 0 & 
  $\Frac{191}{6}$ &   $\Frac{89}{24}$ & 
  $\Frac{118}{3}$ &   $\Frac{269}{24}$ \\ 
0 & 3 & 0 & 0 & 0 & 
  $\Frac{2417}{6}$ &   $\Frac{1141}{48}$ & 
  $573$ &   $\Frac{853}{16}$ \\ 
0 & 4 & 0 & 0 & 0 & 
  $\Frac{2597681}{288}$ &   $\Frac{11465}{36}$ & 
  $\Frac{3916397}{288}$ &   $\Frac{45709}{72}$ \\ 
0 & 5 & 0 & 0 & 0 & 
  $\Frac{19226297}{72}$ &   $\Frac{7313267}{1152}$ & 
  $\Frac{245546135}{576}$ &   $\Frac{14591195}{1152}$ \\ 
\hline
E & L & $gl$ & $q_1$ & $q_2$ & connected & 1PI & 
        conn. (CT) & 1PI (CT) \\ 
\hline
1 & 1 & 1 & 0 & 0 & 
  $\Frac{15}{2}$ &   $\Frac{15}{2}$ & 
  $\Frac{15}{2}$ &   $\Frac{15}{2}$ \\ 
1 & 2 & 1 & 0 & 0 & 
  $\Frac{2383}{24}$ &   $\Frac{67}{6}$ & 
  $\Frac{2923}{24}$ &   $\Frac{157}{6}$ \\ 
1 & 3 & 1 & 0 & 0 & 
  $\Frac{124199}{48}$ &   $\Frac{3533}{24}$ & 
  $\Frac{164807}{48}$ &   $\Frac{6109}{24}$ \\ 
1 & 4 & 1 & 0 & 0 & 
  $\Frac{33696533}{384}$ &   $\Frac{72421}{24}$ & 
  $\Frac{15994703}{128}$ &   $\Frac{123833}{24}$ \\ 
1 & 5 & 1 & 0 & 0 & 
  $\Frac{8024429617}{2304}$ &   $\Frac{5842663}{72}$ & 
  $\Frac{12200014181}{2304}$ &   $\Frac{2605201}{18}$ \\ 
\hline
E & L & $gl$ & $q_1$ & $q_2$ & connected & 1PI & 
        conn. (CT) & 1PI (CT) \\ 
\hline
2 & 1 & 2 & 0 & 0 & 
  $\Frac{31}{2}$ &   $8$ & 
  $\Frac{33}{2}$ &   $9$ \\ 
2 & 2 & 2 & 0 & 0 & 
  $\Frac{1621}{3}$ &   $\Frac{587}{12}$ & 
  $\Frac{1903}{3}$ &   $\Frac{971}{12}$ \\ 
2 & 3 & 2 & 0 & 0 & 
  $\Frac{181157}{8}$ &   $\Frac{4457}{4}$ & 
  $\Frac{692233}{24}$ &   $\Frac{9809}{6}$ \\ 
2 & 4 & 2 & 0 & 0 & 
  $\Frac{33883405}{32}$ &   $\Frac{523703}{16}$ & 
  $\Frac{46451897}{32}$ &   $\Frac{2422249}{48}$ \\ 
2 & 1 & 0 & 2 & 0 & 
  $\Frac{17}{2}$ &   $1$ & 
  $\Frac{19}{2}$ &   $2$ \\ 
2 & 2 & 0 & 2 & 0 & 
  $\Frac{4741}{24}$ &   $11$ & 
  $\Frac{6013}{24}$ &   $16$ \\ 
2 & 3 & 0 & 2 & 0 & 
  $\Frac{101731}{16}$ &   $\Frac{2489}{12}$ & 
  $\Frac{139813}{16}$ &   $\Frac{3821}{12}$ \\ 
2 & 4 & 0 & 2 & 0 & 
  $\Frac{281573453}{1152}$ &   $\Frac{64357}{12}$ & 
  $\Frac{415076813}{1152}$ &   $\Frac{35201}{4}$ \\ 
\hline
E & L & $gl$ & $q_1$ & $q_2$ & connected & 1PI & 
        conn. (CT) & 1PI (CT) \\ 
\hline
3 & 0 & 3 & 0 & 0 & 
  $1$ &   $1$ & 
  $1$ &   $1$ \\ 
3 & 1 & 3 & 0 & 0 & 
  $\Frac{141}{2}$ &   $\Frac{33}{2}$ & 
  $\Frac{149}{2}$ &   $\Frac{35}{2}$ \\ 
3 & 2 & 3 & 0 & 0 & 
  $\Frac{101401}{24}$ &   $\Frac{1097}{4}$ & 
  $\Frac{116437}{24}$ &   $\Frac{1485}{4}$ \\ 
3 & 3 & 3 & 0 & 0 & 
  $\Frac{12090827}{48}$ &   $\Frac{79097}{8}$ & 
  $\Frac{15005863}{48}$ &   $\Frac{105231}{8}$ \\ 
3 & 4 & 3 & 0 & 0 & 
  $\Frac{5869833661}{384}$ &   $\Frac{19132001}{48}$ & 
  $\Frac{7831563637}{384}$ &   $\Frac{27380549}{48}$ \\ 
3 & 0 & 1 & 2 & 0 & 
  $1$ &   $1$ & 
  $1$ &   $1$ \\ 
3 & 1 & 1 & 2 & 0 & 
  $\Frac{69}{2}$ &   $2$ & 
  $\Frac{77}{2}$ &   $3$ \\ 
3 & 2 & 1 & 2 & 0 & 
  $\Frac{8671}{6}$ &   $\Frac{113}{2}$ & 
  $\Frac{5303}{3}$ &   $\Frac{139}{2}$ \\ 
3 & 3 & 1 & 2 & 0 & 
  $\Frac{1617019}{24}$ &   $\Frac{3481}{2}$ & 
  $\Frac{2140925}{24}$ &   $2403$ \\ 
3 & 4 & 1 & 2 & 0 & 
  $\Frac{488989393}{144}$ &   $\Frac{1502905}{24}$ & 
  $\Frac{695346481}{144}$ &   $\Frac{2258617}{24}$ \\ 
\hline
E & L & $gl$ & $q_1$ & $q_2$ & connected & 1PI & 
        conn. (CT) & 1PI (CT) \\ 
\hline
4 & 0 & 4 & 0 & 0 & 
  $4$ &   $1$ & 
  $4$ &   $1$ \\ 
4 & 1 & 4 & 0 & 0 & 
  $491$ &   $\Frac{105}{2}$ & 
  $517$ &   $\Frac{107}{2}$ \\ 
4 & 2 & 4 & 0 & 0 & 
  $\Frac{259367}{6}$ &   $\Frac{7513}{4}$ & 
  $\Frac{293813}{6}$ &   $\Frac{9125}{4}$ \\ 
4 & 3 & 4 & 0 & 0 & 
  $3406344$ &   $\Frac{803681}{8}$ & 
  $\Frac{16607145}{4}$ &   $\Frac{1003895}{8}$ \\ 
4 & 1 & 2 & 2 & 0 & 
  $\Frac{435}{2}$ &   $7$ & 
  $\Frac{477}{2}$ &   $7$ \\ 
4 & 2 & 2 & 2 & 0 & 
  $\Frac{332795}{24}$ &   $\Frac{741}{2}$ & 
  $\Frac{397163}{24}$ &   $\Frac{849}{2}$ \\ 
4 & 3 & 2 & 2 & 0 & 
  $\Frac{13902103}{16}$ &   $\Frac{204421}{12}$ & 
  $\Frac{53753347}{48}$ &   $\Frac{263995}{12}$ \\ 
4 & 1 & 0 & 4 & 0 & 
  $111$ &   $4$ & 
  $125$ &   $4$ \\ 
4 & 2 & 0 & 4 & 0 & 
  $5816$ &   $138$ & 
  $7212$ &   $170$ \\ 
4 & 3 & 0 & 4 & 0 & 
  $\Frac{3747299}{12}$ &   $\Frac{15763}{3}$ & 
  $\Frac{1681643}{4}$ &   $\Frac{21589}{3}$ \\ 
4 & 1 & 0 & 2 & 2 & 
  $\Frac{111}{2}$ &   $2$ & 
  $\Frac{125}{2}$ &   $2$ \\ 
4 & 2 & 0 & 2 & 2 & 
  $2908$ &   $69$ & 
  $3606$ &   $85$ \\ 
4 & 3 & 0 & 2 & 2 & 
  $\Frac{3747299}{24}$ &   $\Frac{15763}{6}$ & 
  $\Frac{1681643}{8}$ &   $\Frac{21589}{6}$ \\ 
\hline
\end{longtable}
}